\documentclass[aps,pra,preprint,groupedaddress]{revtex4}
\usepackage[dvips]{graphicx}
\usepackage{dcolumn} 
\usepackage{bm}      
\usepackage{psfrag}
\begin{document}
\title{On the relationship between parametric two-electron reduced-density-matrix methods and the coupled electron pair approximation}

\author{A. Eugene DePrince III$^{1,2}$ and David A. Mazziotti$^1$}
\affiliation{Department of Chemistry and The James Franck~Institute
\ $^1$ The University of Chicago, Chicago, IL 60637 \\
\ $^2$ Center for Nanoscale Materials, Argonne National Laboratory, Argonne, IL 60439}
\date{August 31, 2010}

\begin{abstract}
Parametric two-electron reduced-density-matrix (p-2RDM) methods have
enjoyed much success in recent years; the methods have been shown to
exhibit accuracies greater than coupled cluster with single and
double substitutions (CCSD) for both closed- and open-shell
ground-state energies, properties, geometric parameters, and
harmonic frequencies.  The class of methods is herein discussed
within the context of the coupled electron pair approximation
(CEPA), and several CEPA-like topological factors are presented for
use within the p-2RDM framework.  The resulting p-2RDM/$n$ methods
can be viewed as a density-based generalization of CEPA/$n$ family
that are numerically very similar to traditional CEPA methodologies.
We cite the important distinction that the obtained energies
represent stationary points, facilitating the efficient evaluation
of properties and geometric derivatives.  The p-2RDM/$n$ formalism
is generalized for an equal treatment of
exclusion-principle-violating (EPV) diagrams that occur in the
occupied and virtual spaces.  One of these general topological
factors is shown to be identical to that proposed by Kollmar [C.
Kollmar, J. Chem. Phys. {\bf 125}, 084108 (2006)], derived in an
effort to approximately enforce the $D$, $Q$, and $G$ conditions for
$N$-representability in his size-extensive density matrix
functional.
\end{abstract}

\maketitle

\section{Introduction}
It has long been understood that the ground-state energy for a many-electron
system can be expressed as a functional of the two-electron
reduced-density-matrix ($2$-RDM) \cite{REF:7_Challenge,REF:7_RDMBook} and thus
the energy may be determined without knowledge of the $N$-electron
wave function.  The direct determination of the $2$-RDM cannot be accomplished
without imposing the so-called $N$-representability conditions
\cite{REF:7_Challenge,REF:7_NREP1,REF:7_NREP2}: constraints that guarantee
that the $2$-RDM corresponds to a realistic $N$-body density.  Two general
classes of methods for the determination of the $2$-RDM without knowledge of the
many electron wave function have emerged. The $2$-RDM may be determined
(i) variationally, whereby the energy is minimized with respect to the elements
of the $2$-RDM under the constraint that the eigenvalues of the matrix remain
non-negative
\cite{REF:7_VAR1,REF:7_VAR2,REF:7_VAR3,REF:7_VAR4,REF:7_VAR5,REF:7_VAR6,REF:7_VAR7},
 or (ii) non-variationally, via the solution of the
anti-Hermitian contracted Schr\"{o}dinger equation (ACSE)
\cite{REF:7_ACSE1,REF:7_ACSE2,REF:7_ACSE3,REF:7_ACSE4,REF:7_ACSE5,REF:7_ACSE6,REF:7_ACSE7,REF:7_ACSE8,REF:7_ACSE9,REF:7_ACSE10,REF:7_ACSE11}.
In the non-variational formalism, the $N$-representability conditions take the
form of the cumulant reconstruction of the $3$-RDM.

The $D$, $Q$, and $G$ conditions for $N$-representability have been
approximately incorporated into a size-extensive functional of the $2$-RDM
as parametrized by a set of single and double excitation coefficients
\cite{REF:7_Kollmar,REF:7_2-RDMF1,REF:7_2-RDMF2,REF:7_2-RDMF3,REF:7_PRL,REF:7_2-RDMF4,REF:7_2-RDMF5,REF:7_2-RDMF6}.
This parametric approach to variational 2-RDM theory exhibits accuracies
superior to CCSD in terms of energies, properties, geometric parameters, and
harmonic frequencies.  The methodology has also been extended to include
general forms for the equivalent treatment of both open- and closed-shell
systems and for use within local correlation approximations suitable for the
treatment of very large molecules \cite{REF:7_2-RDMF5}. Recently, an alternate
parametrization has been developed developed by one of the authors that
stresses the equal treatment of occupied and virtual spaces exhibiting the
accuracy of CCSD with perturbative triple excitations (CCSD(T)) at equilibrium
\cite{REF:7_PRL}, and unlike CCSD(T), the quality of the $2$-RDM solution does
not degrade when stretching a single chemical bond.

The coupled electron pair approximation (CEPA)
\cite{REF:7_CEPA1,REF:7_CEPA2,REF:7_CEPA3,REF:7_CEPA4,REF:7_CEPA5}
and coupled pair functional (CPF) \cite{REF:7_CPF1,REF:7_CPF2}
enjoyed much success in the 1970s, but fell out of use with the advent of
efficient vectorized coupled-cluster algorithms.  By disregarding some diagrams
describing disconnected triple and quadruple excitation amplitudes, coupled
pair theories can be implemented at a cost that is slightly less than CCSD. The
methods are therefore often viewed as at best an approximation to CCSD.
Both early and recent numerical examples, however, have demonstrated that CEPA
is no less accurate for single-reference problems. Although formally less
complete than CCSD, coupled pair methods have many nice properties that have
made them the subject of considerable interest lately
\cite{REF:7_CEPA6,REF:7_CEPA7,REF:7_CEPA8,REF:7_CEPA9,REF:7_CEPA10,REF:7_CEPA11}.
Provided one works within a local orbital basis, CEPA is size-extensive.  CEPA
is conceptually simpler than CCSD, and while formally exhibiting
the same scaling, may be more suitable for large-scale parallelization. It has
recently been demonstrated that the CEPA/1 variant yields thermochemical
properties intermediate in quality between CCSD and CCSD(T) \cite{REF:7_CEPA8}.
Furthermore, the CEPA methods can be expressed within the framework of the CPF
and the topological matrix originally proposed by Ahlrichs \cite{REF:7_CPF1},
implying that the CEPA variants may be incorporated into a CI framework via a
diagonal shift to the Hamiltonian matrix \cite{REF:7_CEPA6,REF:7_CEPA9}. As a
result, efficient CI algorithms can be utilized to solve the CEPA equations.

It is clear that there are connections between the topological matrix of
CI-driven CEPA and the topological factor that lends the parametric $2$-RDM
method its size-extensivity properties. In this paper, we derive CEPA-like
overlap equations for the parametric
$2$-RDM method to elucidate the connection between CEPA and $2$-RDM methods.
In this formalism, the 2-RDM method appears very similar to CEPA with two key
exceptions. First, unlike traditional CEPA methods, parametric $2$-RDM methods
account for the so-called exclusion-principle-violating (EPV) diagrams that
occur in the virtual space; these considerations are a consequence of the
balanced treatment of particles and holes that emerges when considering the~$D$,
~$Q$, and~$G$ conditions for $N$-representability. Second, the overlap
equations for the parametric $2$-RDM method contain an additional term that
renders the solution stationary in all of its variables.  Accordingly, the
determination of density matrices and thus properties and the evaluation of
geometric derivatives are all greatly simplified.  In fact,
the derivative of the energy with respect to an arbitrary nuclear coordinate
requires only the derivatives of the 1- and 2-electron
integrals, which may be evaluated either analytically or numerically.

From explicit relations between the non-variational CEPA and the
variational parametric 2-RDM methods based on the observations
above, we derive topological factors for the parametric 2-RDM
methods that correspond to the CEPA/$n$ variants (with $n=1,2,3$),
as well as new factors that correspond to the same type of
hierarchy, but with a balanced treatment of occupied and virtual
spaces.  The topological factor proposed by Kollmar
\cite{REF:7_Kollmar} is shown to be one that corresponds to variant
$1$ of a CEPA theory that accounts for EPV diagrams in the virtual
space.
We demonstrate the similarities of the two methodologies numerically with
applications to bond stretches and geometry optimizations for several small
molecules.  We also demonstrate the necessity of the
balanced description of particles and holes with a bond stretch for the CH
radical.  In this difficult case, most single-reference theories exhibit
a qualitatively unphysical hump in the potential energy surface; accounting
for EPV diagrams in the virtual space alleviates this problem.  Finally, it is
well known that the CEPA methodologies (with the expectation of the CEPA/0
variant) are not invariant to unitary transformations of the occupied orbitals.
We investigate the numerical behavior of the 2-RDM method with respect to
the same types of orbital rotations.  The 2-RDM method is shown to vary
slightly with the choice of orbitals, but this variance is insignificant
compared to the total correlation energy.  The 2-RDM method is shown to
be rigorously size-extensive for noninteracting two-electron and two-hole
systems when these systems are described by a basis of localized molecular
orbitals.

The key relations of this paper between the non-variational CEPA and
the variational parametric 2-RDM methods were first presented by
DePrince in his Ph.D. thesis at The University of Chicago in
2009~\cite{ED09}. Similar results have appeared in a recent paper,
published to the web on August 18, 2010, by Neese and
Kollmar~\cite{NK10}.  To facilitate circulation of our work, we
publish the present paper to the Archives while we finish a more
complete version of the paper for publication elsewhere.

\section{Theory}
In Section \ref{SECTION:2-RDM} we briefly review parametric $2$-RDM theory.
Section \ref{SECTION:CEPA} outlines the coupled electron pair approximation, and
Section \ref{SECTION:DCEPA} provides a link between the two methodologies.
\subsection{Parametric 2-RDM methods}
\label{SECTION:2-RDM}
To most easily elucidate the connections between parametric $2$-RDM and CEPA
theories, we limit our discussion at this point to the
configuration interaction wave function with only double excitations:
\begin{equation}
| \Psi\rangle = c_0 |\Psi_0\rangle + \mathop{\sum_{a<b}}_{i<j} c_{ij}^{ab} | \Psi_{ij}^{ab} \rangle,
\end{equation}
where $|\Psi_0\rangle$ represents the reference wave function,
$|\Psi_{ij}^{ab}\rangle$ represents a doubly substituted configuration
in which occupied spin-orbitals $i$ and $j$ have been replaced by virtual
spin-orbitals $a$ and $b$, and the set of coefficients
$\{c_0,c_{ij}^{ab}\}$ represents the
respective CI expansion coefficients for these configurations.  For the
normalized wave function, the well-known lack of size-extensivity associated
with truncated CI methods is attributed to those terms in which the excited
determinants interact with the reference configuration.
Size-extensivity may be restored by the introduction of a purely connected
generalized normalization coefficient into the corresponding energy expression,
\begin{equation}
\label{EQN:7_Ec}
E_c = \mathop{\sum_{a<b}}_{i<j} \langle \Psi_0 | \hat{H} | \Psi_{ij}^{ab} \rangle c_{ij}^{ab} c_{0,ij}^{~~ab} + \mathop{\sum_{a<b}}_{i<j} \mathop{\sum_{c<d}}_{k<l} \langle \Psi_{kl}^{cd} | \hat{H} - E_0| \Psi_{ij}^{ab} \rangle c_{kl}^{cd} c_{ij}^{ab},
\end{equation}
where the generalized leading coefficient, $c_{0,ij}^{~~ab}$, is defined
according to Kollmar's definition \cite{REF:7_Kollmar} as
\begin{equation}
\label{EQN:7_Norm}
c_{0,ij}^{~~ab} = (1 - \mathop{\sum_{c<d}}_{k<l} |c_{kl}^{cd}|^2 ~{}^2f_{ijkl}^{abcd} )^{1/2}.
\end{equation}
The $8$-index topological matrix, ${}^2f_{ijkl}^{abcd}$, interpolates between
$N$-representable the (but not size-extensive)
CID solution, recovered by setting all ${}^2f_{ijkl}^{abcd}$ equal to unity,
and the size-extensive (but not $N$-representable) CEPA/0 solution, obtained
by setting all ${}^2f_{ijkl}^{abcd}$ equal to zero.  It is true that the CEPA/0
parametrization will restore size-extensivity, but we arrive at a more
intelligent choice by recognizing that the $N$-representability of the
associated $2$-RDM is strictly dependent upon the form of the topological
factor.  We will choose this factor such that it will enforce, at least
approximately, the known two-particle $N$-representability conditions, the
so-called~$D$,~$Q$, and~$G$ conditions. With these considerations,
Kollmar proposed in Ref. \cite{REF:7_Kollmar} the following factor:
\begin{eqnarray}
f_{ijkl}^{abcd} = F_{ij}^{kl} + F_{ab}^{cd} - F_{ij}^{kl}F_{ab}^{cd}, \\
F_{pq}^{st} = \frac{1}{4} ( \delta_{ps} + \delta_{pt} + \delta_{qs} + \delta_{qt}),
\end{eqnarray}
where the delta function $\delta_{pq}$ is zero when the spatial component of
orbitals $p$ and $q$ are disjoint and one otherwise.  By replacing
$c_{0,ij}^{~~ab}$ in Eq. (\ref{EQN:7_Ec}) with its definition in Eq. (\ref{EQN:7_Norm}), the correlation energy may be determined via an unconstrained minimization.

\subsection{The coupled electron pair approximation}
\label{SECTION:CEPA}
Beginning with the CID energy functional, a set of non-linear equations may be
obtained by enforcing the stationary condition, $\partial E_c/\partial c_{ij}^{ab} = 0$, to obtain
\begin{eqnarray}
E_c = \mathop{\sum_{a<b}}_{i<j}\frac{c_{ij}^{ab}}{c_0}\langle\Psi_{ij}^{ab}|\hat{H}|\Psi_0\rangle, \\
c_{ij}^{ab} E_c = c_0 \langle \Psi_{ij}^{ab} |\hat{H}|\Psi_0\rangle + \mathop{\sum_{c<d}}_{k<l} \langle\Psi_{ij}^{ab} |\hat{H}-E_0|\Psi_{kl}^{cd}\rangle c_{kl}^{cd}.
\end{eqnarray}
Making a change of variables, $b_{ij}^{ab} = c_{ij}^{ab}/c_0$, yields the
CID overlap equations in intermediate normalization:
\begin{eqnarray}
\label{EQN:7_CID_E}
E_c = \mathop{\sum_{a<b}}_{i<j}b_{ij}^{ab}\langle\Psi_{ij}^{ab}|\hat{H}|\Psi_0\rangle, \\
\label{EQN:7_CID_COEF}
E_c b_{ij}^{ab} = \langle \Psi_{ij}^{ab} |\hat{H}|\Psi_0\rangle + \mathop{\sum_{c<d}}_{k<l} \langle\Psi_{ij}^{ab} |\hat{H}-E_0 |\Psi_{kl}^{cd}\rangle b_{kl}^{cd}.
\end{eqnarray}
The left hand side of Eq. (\ref{EQN:7_CID_COEF}) is clearly quadratic; the
size-extensivity problem in this representation amounts to the lack of a
complementary quadratic term
on the right hand side of this equation.  Such a term describes the interaction
between all doubly and quadruply excited configurations:
\begin{equation}
\label{EQN:7_cidq}
0 = \langle \Psi_{ij}^{ab} |\hat{H}|\Psi_0\rangle + \mathop{\sum_{c<d}}_{k<l}
\langle \Psi_{ij}^{ab} | \hat{H} - E_0 - E_c| \Psi_{kl}^{cd}\rangle b_{kl}^{cd} + \langle \Psi_{ij}^{ab} | \hat{H} | \Psi_Q \rangle,
\end{equation}
where $|\Psi_Q\rangle$ includes all quadruply excited configurations,
$|\Psi_{ijkl}^{abcd}\rangle$, and their respective intermediately normalized
CI expansion coefficients, $b_{ijkl}^{abcd}$.  The
coupled-cluster with doubles (CCD) equations are recovered by approximating
the coefficients of the quadruples, $b_{ijkl}^{abcd}$ as an antisymmetric
sum of products of doubles coefficients as suggested by second order
perturbation theory.  The CEPA methods imply a simpler
relationship between the double and quadruple coefficients by taking only the
leading term of this sum:
\begin{equation}
\label{EQN:7_c4}
b_{ijkl}^{abcd} = b_{ij}^{ab} b_{kl}^{cd}.
\end{equation}
Inserting Eq. (\ref{EQN:7_c4}) into Eq. (\ref{EQN:7_cidq}) yields
\begin{equation}
0 = \langle \Psi_{ij}^{ab} |\hat{H}|\Psi_0\rangle + \mathop{\sum_{c<d}}_{k<l} \langle \Psi_{ij}^{ab} | \hat{H}-E_0 - E_c | \Psi_{kl}^{cd}\rangle b_{kl}^{cd} + \mathop{\sum_{c<d}}_{k<l}\langle \Psi_{ij}^{ab} | \hat{H} | \Psi_{ijkl}^{abcd} \rangle b_{ij}^{ab}b_{kl}^{cd},
\end{equation}
and the last term may be reexpressed equivalently according to Slater's rules
to give
\begin{equation}
0 = \langle \Psi_{ij}^{ab} |\hat{H}|\Psi_0\rangle + \mathop{\sum_{c<d}}_{k<l} \langle \Psi_{ij}^{ab} | \hat{H}-E_0-E_c | \Psi_{kl}^{cd}\rangle b_{kl}^{cd} + \mathop{\sum_{c<d}}_{k<l}\langle \Psi_{0} | \hat{H} | \Psi_{kl}^{cd} \rangle b_{ij}^{ab}b_{kl}^{cd},
\end{equation}
or
\begin{equation}
0 = \langle \Psi_{ij}^{ab} |\hat{H}|\Psi_0\rangle + \mathop{\sum_{c<d}}_{k<l} \langle \Psi_{ij}^{ab} | \hat{H}-E_0-E_c | \Psi_{kl}^{cd}\rangle b_{kl}^{cd} + E_c b_{ij}^{ab}.
\end{equation}
We have arrived at the simplest CEPA approximation, denoted CEPA/0:
\begin{equation}
\label{EQN:7_cepa0}
0 = \langle \Psi_{ij}^{ab} |\hat{H}|\Psi_0\rangle + \mathop{\sum_{c<d}}_{k<l} \langle \Psi_{ij}^{ab} | \hat{H}-E_0 | \Psi_{kl}^{cd}\rangle b_{kl}^{cd}.
\end{equation}
The CEPA/0 approximation is a naive one in that we have unintentionally included
the effects of a large number of unphysical terms.  Equation (\ref{EQN:7_cepa0})
deteriorates whenever the indices $\{ij\}$ and $\{kl\}$
(or $\{ab\}$ and $\{cd\}$)
have any coincidences.  Such instances imply multiple excitations out of
(or into) the same orbitals twice and are therefore referred to as
exclusion-principle-violating (EPV) terms.  The remainder of the CEPA
approximations differ only in how they account for EPV terms.

By defining a diagonal shift, $\Delta_{ij}^{ab}$, we may write general
equations for all of the CEPA variants that improve upon CEPA/0,
\begin{equation}
\label{EQN:7_SHIFT}
0 = \langle \Psi_{ij}^{ab} |\hat{H}|\Psi_0\rangle + \mathop{\sum_{c<d}}_{k<l} \langle \Psi_{ij}^{ab} | \hat{H}-E_0-\Delta_{ij}^{ab} | \Psi_{kl}^{cd}\rangle b_{kl}^{cd},
\end{equation}
where we can easily see that $\Delta_{ij}^{ab}$ is equal to $-E_c$ for CID and
zero for CEPA/0. The simplest
improvement upon CEPA/0, called CEPA/2, removes only those EPV terms of the form
$b_{ij}^{ab}b_{ij}^{cd}$; the diagonal shift for CEPA/2 is defined as the pair
energy, $e_{ij}$,
\begin{equation}
\label{EQN:7_eij}
\Delta_{ij}^{ab} = e_{ij} = \sum_{c<d} b_{ij}^{cd} \langle \Psi_{ij}^{cd} | \hat{H} | \Psi_0 \rangle.
\end{equation}
Table \ref{TAB:7_1} lists the definitions of the diagonal shifts for the CID
and CEPA/$n$ ($n=0,1,2,3$) methods.
\begin{table}[htpb!]
  \caption{Diagonal shifts that define the CID and CEPA equations.  The various
           CEPA shifts lend  size-extensivity to CID
           while removing varying degrees of unphysical exclusion principle
           violating (EPV) terms from the overlap equations.}
  \label{TAB:7_1}
  \begin{center}
  \begin{ruledtabular}
    \begin{tabular}{cc}
      Method & $\Delta_{ij}^{ab}$ \\
      \hline
      CID &  $-E_c$ \\
      CEPA/0 &  0 \\
      CEPA/1 &  $\frac{1}{2}\sum_k(e_{ik}+e_{jk})$ \\
      CEPA/2 &  $e_{ij}$ \\
      CEPA/3 &  $\sum_k(e_{ik}+e_{jk})-e_{ij}$\\
    \end{tabular}
    \end{ruledtabular}
  \end{center}
\end{table}
CEPA/3 removes all EPV diagrams in the
occupied space, and the CEPA/1 shift can be viewed as the average of those
corresponding to CEPA/2 and CEPA/3.  In general, the number of EPV diagrams
removed by each flavor of CEPA is CEPA/3 $>$ CEPA/1 $>$ CEPA/2 $>$ CEPA/0, and
as such the correlation energy is lowest for CEPA/0 and highest for CEPA/3.
We may express $e_{ij}$ in a more suggestive form by incorporating idea of the
topological factor into Eq. (\ref{EQN:7_eij}), and symmetrizing the factor with
respect to the exchange of orbitals $i$ and $j$ or $k$ and $l$,
\begin{equation}
\Delta_{ij}^{ab} = \mathop{\sum_{c<d}}_{k<l}\langle\Psi_{kl}^{cd}\hat{H}|\Psi_0\rangle b_{kl}^{cd} ~{}^2f_{ijkl}^{abcd},
\end{equation}
where
\begin{equation}
{}^2f_{ijkl}^{abcd} = \frac{1}{2}(\delta_{ik}\delta_{jl}+\delta_{jk}\delta_{il}).
\end{equation}
It is immediately clear that one could define a topological factor corresponding
for each of the CEPA/$n$ variations;  these factors, symmetrized with respect
to orbital exchange, are presented in Table \ref{TAB:7_2}.
\begin{table}[htpb!]
\caption
        {Symmetrized topological factors corresponding to the
        CEPA/$n$ family of methods.  Note that EPV diagrams are
        neglected in virtual space.  When incorporated into the
        parametric 2-RDM method, the methods are denoted p-2RDM/$n$.}
  \label{TAB:7_2}
  \begin{center}
    \begin{ruledtabular}
    \begin{tabular}{cc}
      Method & $f_{ijkl}^{abcd}$ \\
      \hline
      CID &  1  \\
      CEPA/0 &  0 \\
      CEPA/1 & $\frac{1}{4}(\delta_{ik}+\delta_{jl}+\delta_{il}+\delta_{jk})$ \\
      CEPA/2 & $\frac{1}{2}(\delta_{ik}\delta_{jl}+\delta_{il}\delta_{jk})$\\
      CEPA/3 & $\frac{1}{2}(\delta_{ik}+\delta_{jl}+\delta_{il}+\delta_{jk} - (\delta_{ik}\delta_{jl}+\delta_{il}\delta_{jk}))$ \\
    \end{tabular}
    \end{ruledtabular}
  \end{center}
\end{table}
We will show in the next
section that these factors can be incorporated into the parametric 2-RDM
energy functional to yield a family of density-based CEPA-like methods, which
we call p-2RDM/$n$.

\subsection{Density-based CEPA}
\label{SECTION:DCEPA}
In this section, we illustrate the connection between the parametric $2$-RDM
method and the CEPA/$n$ family of equations.
By enforcing the stationary condition on Eq. (\ref{EQN:7_Ec}), we obtain the
following system of coupled nonlinear equations that define the parametric
2-RDM energy and excitation coefficients:
\begin{eqnarray}
\label{EQN:7_Ecorr3}
E_c = \mathop{\sum_{a<b}}_{i<j}\frac{c_{ij}^{ab}}{c_{0,ij}^{~~ab}} \langle \Psi_0 | \hat{H} | \Psi_{ij}^{ab} \rangle, \\
0 = \langle \Psi_{kl}^{cd} |\hat{H}|\Psi_0\rangle  c_{0,kl}^{cd}
- c_{kl}^{cd} \mathop{\sum_{a<b}}_{i<j}  ~{}^2f_{ijkl}^{abcd}\langle \Psi_0 |\hat{H}|\Psi_{ij}^{ab}\rangle  \frac{c_{ij}^{ab}}{c_{0,ij}^{~~ab}}
+ \mathop{\sum_{a<b}}_{i<j}\langle \Psi_{kl}^{cd} | \hat{H} -E_0| \Psi_{ij}^{ab} \rangle c_{ij}^{ab}.
\end{eqnarray}
We can make the transformation $b_{ij}^{ab} = c_{ij}^{ab}/c_{0,ij}^{~~ab}$ to
obtain a set of equations that is identical to the CEPA family of equations
given in Eq. (\ref{EQN:7_SHIFT}) with a diagonal shift that is defined as
\begin{eqnarray}
\label{EQN:7_delta}
\Delta_{ij}^{ab} = \mathop{\sum_{c<d}}_{k<l} b_{kl}^{cd} ~{}^2f_{ijkl}^{abcd} \langle \Psi_{kl}^{cd} | \hat{H} | \Psi_0 \rangle + \mathop{\sum_{c<d}}_{k<l}\Bigg ( \frac{c_{0,kl}^{cd}}{c_{0,ij}^{~~ab}}-1\Bigg )\frac{b_{kl}^{cd}}{b_{ij}^{ab}}\langle \Psi_{ij}^{ab} |\hat{H}|\Psi_{kl}^{cd}\rangle.
\end{eqnarray}
Again, from the perspective of coupled pair theories, $\Delta_{ij}^{ab}$
 can be interpreted as an
approximation of the effects of higher excitations neglected in the CI
expansion.  The second term in Eq. (\ref{EQN:7_delta}) can be seen as one that
renders the energy and true minimum and thus that the corresponding density
matrix is a stationary solution to Eq. (\ref{EQN:7_Ec}).  Assuming that this
term is sufficiently small and may be ignored, the connection between
parametric $2$-RDM methods and the CEPA approximations is obvious.  This
assumption is not unreasonable for well-behaved systems; the term is exactly
zero in the both the CID and CEPA/0 limits.  By replacing the topological
factor in Eq. (\ref{EQN:7_Ec}) or (\ref{EQN:7_delta}) with those defined in
Table \ref{TAB:7_2}, we obtain a family of methods with stationary solutions
that yield results that are numerically very similar to traditional CEPA/$n$
implementations.  We term this family of density-based CEPA-like methods
parametric 2-RDM/$n$ or p-2RDM/$n$ methods.

Returning to the original formulation of the parametric $2$-RDM method, the
Kollmar topological factor, which we will call K, can be viewed as
one that is very similar to CEPA/1 with the important distinction that it
accounts for EPV diagrams in the virtual space.  This balanced description
of occupied and virtual spaces is central to reduced-density-matrix theory
(consider the complementary $D$, $Q$, and $G$ conditions for
$N$-representability).  We may define a family of density-based CEPA-like
topological factors to be incorporated into the parametric $2$-RDM formalism
by (i) accounting for EPV diagrams in the virtual space and (ii)
symmetrizing each factor with respect to the exchange of orbitals $i$ and $j$,
$k$ and $l$, $a$ and $b$, or $c$ and $d$.  These factors are defined in Table
\ref{TAB:7_3}.  These new topological matrices yield a family of improved density-based
\begin{table}[htpb!]
  \caption
  {Symmetrized topological factors with a balanced
  description of the occupied and virtual spaces.  Each
  factor, with the exception of CEPA/0, yields the exact
  correlation energy for two-electron systems. Each topological factor
  is defined as a combination of tensors corresponding to the
  occupied and virtual spaces:
  $f_{ijkl}^{abcd}=F_{ij}^{kl}+F_{ab}^{cd} - F_{ij}^{kl}F_{ab}^{cd}$.}
  \label{TAB:7_3}
  \begin{center}
    \begin{ruledtabular}
    \begin{tabular}{cc}
      Method & $F_{pq}^{st}$\\
      \hline
      CID &  1  \\
      CEPA/0 &  0 \\
      p-2RDM$^\prime$/1 & $\frac{1}{4}(\delta_{ps}+\delta_{pt}+\delta_{qs}+\delta_{qt})$\\
      p-2RDM$^\prime$/2 & $\frac{1}{2}(\delta_{ps}\delta_{qt}+\delta_{pt}\delta_{qs})$\\
      p-2RDM$^\prime$/3 & $\frac{1}{2}(\delta_{ps}+\delta_{pt}+\delta_{qs}+\delta_{qt} - (\delta_{ps}\delta_{qt}+\delta_{pt}\delta_{qs}))$ \\
      K & $\frac{1}{4}(\delta_{ps}+\delta_{pt}+\delta_{qs}+\delta_{qt})$ \\
    \end{tabular}
    \end{ruledtabular}
  \end{center}
\end{table}
CEPA methods and are labeled p-2RDM$^\prime$/$n$ (for $n=1,2,3$).
Each p-2RDM$^\prime$/$n$ variant may be implemented at a cost that is
comparable to other two-electron theories.
It should be noted that each factor, with the exception of CEPA/0,
recovers the exact correlation energy in the two-particle (or two-hole) limit.
Furthermore, when incorporated in the parametric $2$-RDM formalism, which is
Hermitian, geometry optimizations, the determination of density matrices, and
the evaluation of one- and two-electron properties are all greatly simplified.

To this point we have not considered the effects of single excitations on
size-extensivity and EPV diagrams.  No clear consensus for the treatment of
single excitations in coupled-pair formalisms is present in the literature.
For this reason, we choose to treat single excitations in each of the
p-2RDM/$n$ and p-2RDM$^\prime$/$n$ methods as is
described in Ref. \cite{REF:7_2-RDMF4}.  We have the generalized
normalization coefficient, $c_{0,ij}^{~~ab}$, defined as
\begin{equation}
c_{0,ij}^{~~ab} = (1 - \mathop{\sum_c}_k|c_{k}^{c}|^2~{}^1f_{ijkk}^{abcc} - \mathop{\sum_{c<d}}_{k<l} |c_{kl}^{cd}|^2 ~{}^2f_{ijkl}^{abcd} )^{1/2},
\end{equation}
where ${}^1f_{ijkk}^{abcc}$ is either defined as
\begin{equation}
\label{EQN:7_f1}
{}^1f_{ijkk}^{abcc} = 1 - (1-\delta_{ik})(1-\delta_{jk}),
\end{equation}
in the p-2RDM/$n$ formalism, or
\begin{equation}
\label{EQN:7_f1v}
{}^1f_{ijkk}^{abcc} = 1 - (1-\delta_{ik})(1-\delta_{jk})(1-\delta_{ac})(1-\delta_{bc}),
\end{equation}
in the improved p-2RDM$^\prime$/$n$ formalism.  The value of
${}^1f_{ijkk}^{abcc}$ as given by Eq. (\ref{EQN:7_f1}) is unity unless the
spatial component of the occupied orbitals $i$ and $j$ are disjoint with $k$;
this is the treatment chosen to coincide the best with existing CEPA theories.
The value of ${}^1f_{ijkk}^{abcc}$ as given by Eq. (\ref{EQN:7_f1}) is unity
unless the spatial components of the occupied orbitals $i$ and $j$ are disjoint
with $k$ {\em and} the spatial components of the virtual orbitals $a$ and $b$
are disjoint with $c$.  From the perspective of coupled pair theories,
our singles topological factors approximate the inclusion of disconnected
triple excitations while removing EPV diagrams in the occupied space or
occupied and virtual spaces.  We note that this treatment of single excitations
removes {\em all} EPV diagrams in either the occupied space or occupied and
virtual spaces arising from single excitations and is thus most similar to the
CEPA/3 treatment of single excitations.

\section{Discussion}
CEPA/$n$ calculations were performed using the Molpro electronic structure
package \cite{REF:7_MOLPRO}.  The closed-shell p-2RDM/$n$ calculations were
performed using our code implemented within
the {\small PSI3} {\em ab initio} electronic structure package
\cite{REF:7_2-RDMF6,REF:7_PSI3}. All
open-shell parametric 2-RDM calculations were performed with a separate code
with all $1$- and $2$-electron integrals obtained from the {\small GAMESS}
electronic structure package \cite{REF:7_GAMESS}.
The topological factors for each p-2RDM/$n$ and p-2RDM$^\prime$/$n$ variant are
presented in Tables \ref{TAB:7_2} and \ref{TAB:7_3}, respectively; note
that, as in the Molpro implementation of CEPA/$n$, EPV diagrams in the virtual
space are neglected for p-2RDM/$n$ calculations. Single excitations are treated
in the p-2RDM/$n$ and p-2RDM$^\prime$/$n$ methods as described in Ref.
\cite{REF:7_2-RDMF4}.

\begin{figure}[!htpb]
  \caption{Potential energy curve for a single O-H bond stretch for H$_2$O
           in a cc-pVDZ basis set with
           one core orbital frozen.  Curves are presented for the CEPA and
           p-2RDM variants 1, 2, and 3.}
  \label{FIG:7_1}
  \begin{center}
    \resizebox{10cm}{!}{\includegraphics{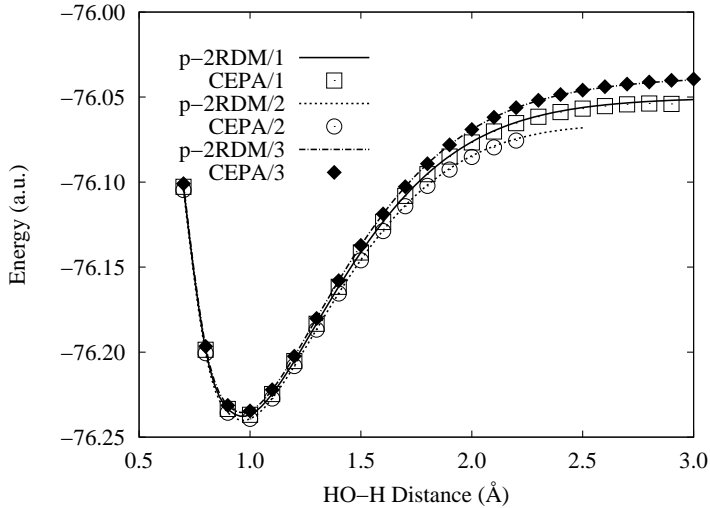}}
  \end{center}
\end{figure}

\begin{figure}[!htpb]
  \caption{Potential energy curve for a single N-H bond stretch for
           NH$_3$ in a cc-pVDZ basis set with
           one core orbital frozen.  The bond length for one hydrogen is
           increased while holding all other bonds and angles constant.
           Curves are presented for the CEPA and p-2RDM variants 1, 2, and 3.}
  \label{FIG:7_2}
  \begin{center}
    \resizebox{10cm}{!}{\includegraphics{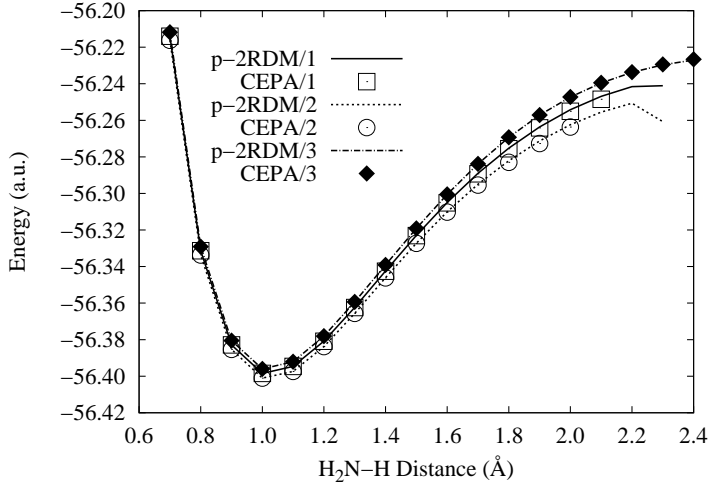}}
  \end{center}
\end{figure}

Figure \ref{FIG:7_1}~illustrates the CEPA/$n$ and p-2RDM/$n$ potential energy
surfaces for a single O-H bon stretch for H$_2$O in a cc-pVDZ basis set.
Clearly the p-2RDM/$n$ and
CEPA/$n$ methods perform identically in all regions of the potential energy
curve. The largest deviation between the two families is 0.95
milli-Hartrees (mH), occurring for p-2RDM/$2$ at a bond length of 2.2 \AA.
Molpro CEPA/2 calculations did not converge beyond 2.2 \AA.
The largest discrepancy with CEPA/$3$ is only 0.47 mH, occurring at 3.0 \AA.
Figure \ref{FIG:7_2}~illustrates similar trends for the NH$_2$-H
bond stretch.  CEPA/$n$ and p-2RDM/$n$ results are indistinguishable over the
range of reported bond lengths.

The p-2RDM/$n$ and CEPA/$n$ formalisms were also applied geometry optimizations
and harmonic frequency analysis for H$_2$O and CO$_2$ in a cc-pVDZ basis set.
The optimized bond lengths and energies for CO$_2$ as computed by the CEPA/$n$
and p-2RDM/$n$ methods are listed in Table \ref{TAB:7_4};
\begin{table}[htpb!]
  \caption
          {Optimized geometries and energies for CO$_2$ in
           a cc-pVDZ basis set. Energies and bond lengths are given
           in Hartrees and \AA, respectively. CEPA/$n$ and p-2RDM/$n$ methods
           yield nearly identical results for both the bond length
           and the minimum energy.  Core orbitals are restricted to
           be occupied.}
  \label{TAB:7_4}
  \begin{center}
    \begin{ruledtabular}
    \begin{tabular}{ccccc}
       & \multicolumn{2}{c}{CEPA} & \multicolumn{2}{c}{p-2RDM} \\
      \cline{2-3} \cline{4-5}
       Variant & Energy  & r$_{\rm CO}$& Energy  & r$_{\rm CO}$ \\
      \hline
      1 & -188.1347 & 1.1713 & -188.1341 & 1.1707 \\
      2 & -188.1440 & 1.1743 & -188.1424 & 1.1729 \\
      3 & -188.1263 & 1.1688 & -188.1263 & 1.1688 \\
    \end{tabular}
    \end{ruledtabular}
  \end{center}
\end{table}
in general we can see that the optimal bond length contracts with a more
rigorous treatment of EPV diagrams.  The CEPA/3 and p-2RDM/3 results in Table
\ref{TAB:7_4} are indistinguishable.  The discrepancies that arise between the
variant 2  and 1 numbers are due to the difference in the treatment of single
excitations in CEPA/$n$ and p-2RDM/$n$ theories.  The p-2RDM/$n$ methods remove
{\em all} EPV diagrams in the occupied space due to single excitations whereas
CEPA/3 removes more than CEPA/1, which removes more than CEPA/2.  Accordingly,
we observe the greatest disparities between the variant 2 values.
Tables \ref{TAB:7_6} and \ref{TAB:7_7} present the optimized energies, geometric
\begin{table}[htpb!]
  \caption{Optimized geometries and energies for H$_2$O in
           a cc-pVDZ basis set. Energies, bond lengths,
           and angles are given in Hartrees, \AA, and degrees,
           respectively. CEPA/$n$ and p-2RDM/$n$ methods yield identical
           results for the geometric parameters and the minimum
           energy.  Core orbitals are restricted to be occupied.
           The experimentally obtained values for r$_{\rm OH}$ and
           a$_{\rm HOH}$ are 0.9578 \AA~and 104.4776 degrees, respectively.}
  \label{TAB:7_6}
  \begin{center}
    \begin{ruledtabular}
    \begin{tabular}{ccccccc}
     & \multicolumn{3}{c}{CEPA} & \multicolumn{3}{c}{p-2RDM} \\
    \cline{2-4} \cline{5-7}
     Variant & Energy  & r$_{\rm OH}$ & a$_{\rm HOH}$ & Energy  & r$_{\rm OH}$ & a$_{\rm HOH}$ \\
    \hline
    1 & -76.2382 & 0.9652 & 102.11 & -76.2382 & 0.9652 & 102.12 \\
    2 & -76.2406 & 0.9663 & 101.99 & -76.2406 & 0.9663 & 102.00 \\
    3 & -76.2360 & 0.9642 & 102.22 & -76.2360 & 0.9642 & 102.22 \\
    \end{tabular}
    \end{ruledtabular}
  \end{center}
\end{table}
\begin{table}[htpb!]
  \caption
{Harmonic frequencies in wavenumbers, cm$^{-1}$, for
          H$_2$O in a cc-pVDZ basis set.  CEPA/$n$ and p-2RMD/$n$ methods
          yield are nearly indistinguishable, with the largest
          discrepancies being 1.9 cm$^{-1}$ for the symmetric and
          asymmetric stretches as described by variant 2.}
  \label{TAB:7_7}
  \begin{center}
    \begin{ruledtabular}
    \begin{tabular}{ccccccc}
       & \multicolumn{3}{c}{CEPA} & \multicolumn{3}{c}{p-2RDM} \\
      \cline{2-4} \cline{5-7}
       Variant & a$_{1}$  & a$_1$ &  b$_{2}$ & a$_{1}$  & a$_1$ &  b$_{2}$ \\
      \hline
      1 & 3837.1 & 1695.3 & 3937.8 & 3838.2 & 1696.4 & 3938.7 \\
      2 & 3815.9 & 1691.8 & 3917.8 & 3817.8 & 1692.9 & 3919.7 \\
      3 & 3855.9 & 1698.5 & 3955.3 & 3855.5 & 1699.4 & 3955.5 \\
    \end{tabular}
    \end{ruledtabular}
  \end{center}
\end{table}
parameters, and harmonic frequencies as computed by the CEPA/$n$ and p-2RDM/$n$
methods for the H$_2$O molecule.  Geometric parameters are identical within
each variant, with a difference in bond angle of only 0.01 degrees for variants
1 and 2.  As was observed with CO$_2$, bond lengths slightly contract with a
more rigorous treatment of EPV diagrams.  Harmonic
frequencies are nearly identical between the two methods, with the largest
difference between CEPA and p-2RDM being less than two wavenumbers.
That larger differences exist between geometric parameters determined by
CEPA and p-2RDM methods in the case of CO$_2$ rather than H$_2$O is
not a surprise.  The treatment of single excitations is directly connected to
the description of disconnected triple excitations, and the
emergence of discrepancies between the two methods is most likely due to the
growing importance of triple excitations for CO$_2$ as compared to H$_2$O.

\begin{figure}[htpb!]
  \caption{Potential energy curve for the bond stretch for the CH radical
           in a cc-pVDZ basis set with
           one core orbital frozen.  CCSD and p-2RDM/3 both exhibit
           and unphysical hump at long bond lengths. p-2RDM$^\prime$/3
           has no such feature.}
  \label{FIG:7_3}
  \begin{center}
    \resizebox{10cm}{!}{\includegraphics{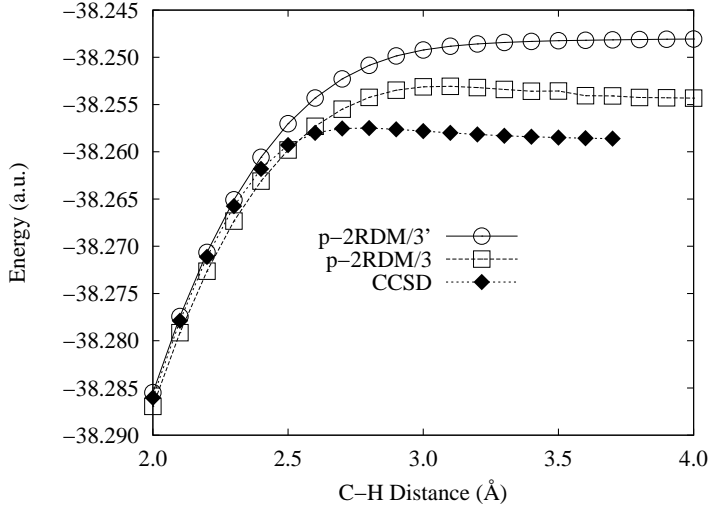}}
  \end{center}
\end{figure}

We next investigate the importance of the equal treatment of EPV diagrams
arising in the occupied and virtual spaces.  In CEPA methodologies, the EPV
terms for the virtual space are generally ignored because they are far fewer
in number than those occurring in the occupied space.  This omission is
simply intended to increase computational efficiency.  There are certain
situations, however, where the neglect of virtual space EPV terms may cause the
CEPA methods to qualitatively fail.  Figure \ref{FIG:7_3} illustrates such a
point.  We
calculated the potential energy curve for the CH radical with CCSD and the
p-2RDM/3 and p-2RDM$^\prime$/3 methods using the topological factors given in
Tables \ref{TAB:7_2} and \ref{TAB:7_3}, respectively. At around 2.8 \AA~an
unphysical hump occurs in the CCSD curve. p-2RDM/3 neglects the virtual space
EPV diagrams and as a result develops a similar hump around 3.1 \AA.  The
p-2RDM$^\prime$/3 method never displays any unphysical behavior.

\begin{figure}[!htpb]
  \label{FIG:INVAR}
  \caption{Energy per helium atom as a function of the number of He atoms
           for a system of non-interacting helium atoms.  The 2-RDM method is
           numerically size-extensive utilizing both canonical and localized
           molecular orbitals.  Size-extensivity is absolutely rigorous in
           the local orbital basis.}
  \begin{center}
    \resizebox{11cm}{!}{\includegraphics{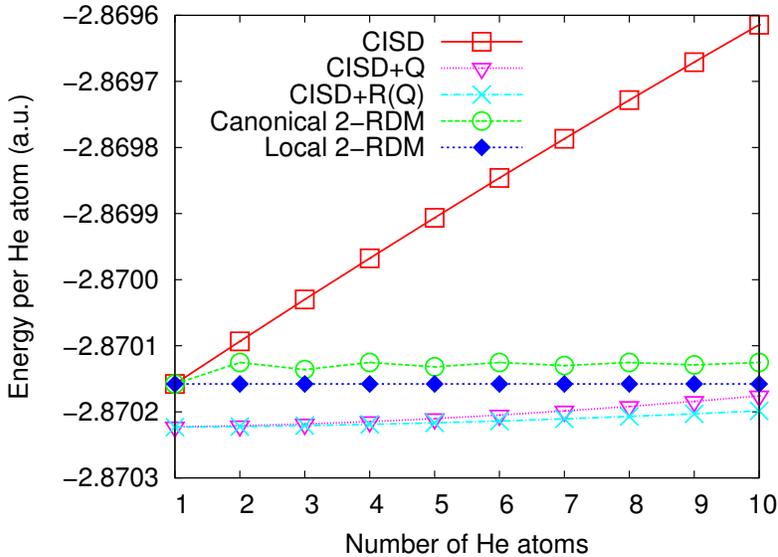}}
  \end{center}
\end{figure}

It is well known that the pair energies associated with the EPV diagrams are
not invariant to rotations among the occupied orbitals \cite{REF:7_BART_INVAR}.
This variance is a consequence of the partial nature of the summations that
define pair energies; the indices involved to not span the entirety of the
Hilbert space, and the pair contribution to the energy is thus not invariant
to unitary transformations.  Accordingly, size-extensivity can only rigorously
be achieved in these coupled-pair theories with the use of a localized
molecular orbital basis.  The close relationship of the parametric 2-RDM method
to these theories suggests that similar deficiencies may exist within the
present formulation of the 2-RDM method.
The numerical size-extensivity of the parametric 2-RDM method was demonstrated
by the authors for a series of infinitely separated He atoms in Ref.
\cite{REF:7_2-RDMF1}. We revisit this system, illustrating in Fig.
\ref{FIG:INVAR} the  size-extensivity and orbital invariance properties (or
lack thereof) for the parametric 2-RDM method. We choose the Kollmar
parametrization, K (also denoted p-2RDM$^\prime$/1 presently), as the
representative example.
We treat an increasing number of helium atoms at effectively infinite separation
to illustrate the size-extensivity of the 2-RDM method in a basis of canonical
and localized orbitals.  The He atoms are situated on a line with an
interatomic distance of 200 \AA; the atomic orbitals are represented by an
Ahlrichs double-zeta basis set \cite{REF:7_ADZ}.  For the localized-orbital
calculations, occupied and virtual orbitals were localized separately
according to the Boys localization criterion \cite{REF:7_Boys}. The
energy per He atom for a size-extensive method should not vary with the number
of He atoms.  We see that the 2-RDM does display this characteristic when
utilizing either localized and delocalized canonical orbitals; regardless of the
``bumps" in the canonical basis, the 2-RDM results do not display any systematic
dependence upon system size. The energies obtained in the local basis, however,
are rigorously size-extensive, with no numerical deviation in the energy per
He atom for all system sizes.  We have also presented the energies obtained from
the Davidson correction \cite{REF:7_Davidson}, denoted CISD+Q, and the
renormalized Davidson correction \cite{REF:7_RD1,REF:7_RD2}, denoted CISD+R(Q).
Clearly, neither the Davidson corrected nor renormalized Davidson corrected
energies are rigorously size-extensive, with both results exhibiting a clear
and systematic dependence on system size.

\begin{figure}[!htpb]
  \caption
  {The percentage change in the correlation energy,
  $\frac{E_{c,\rm{canon}}-E_{c,\rm{local}}}{E_{c,\rm{canon}}}\times 100\%$
  when utilizing localized molecular orbitals rather than canonical
  Hartree-Fock orbitals for the H-F bond stretch in (a) cc-pVDZ and (b)
  cc-pVTZ basis sets.}
  \label{FIG:INVAR2}
  \begin{center}
    \resizebox{11cm}{!}{\includegraphics{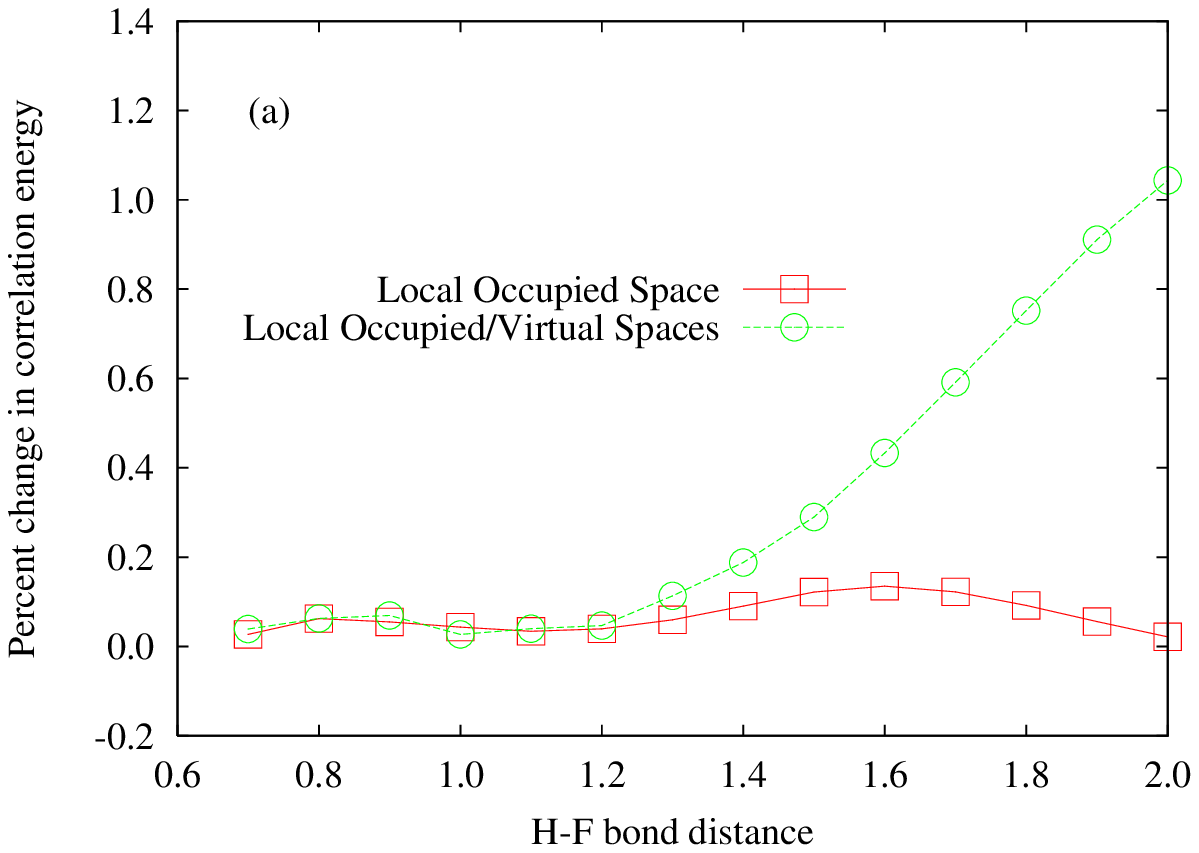}}
    \resizebox{11cm}{!}{\includegraphics{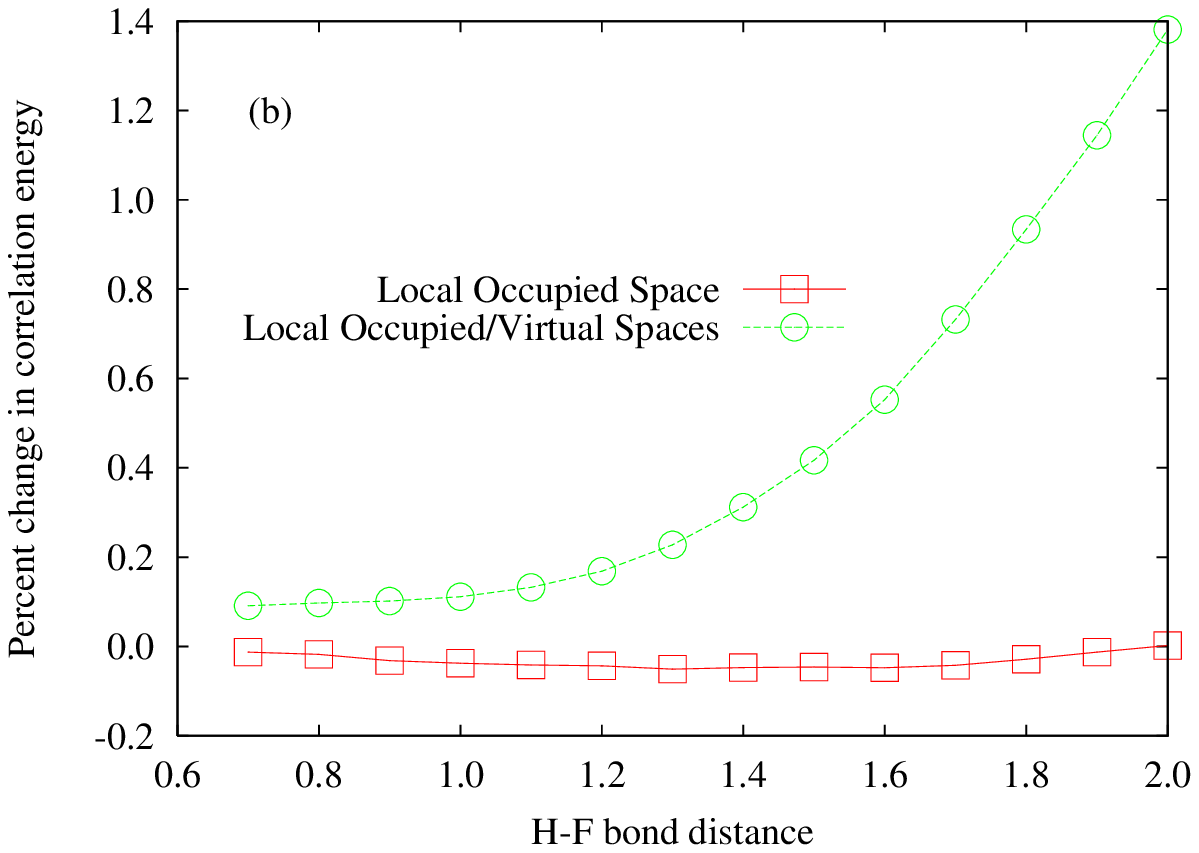}}
  \end{center}
\end{figure}

These results unfortunately demonstrate that the 2-RDM energy is not
strictly invariant to unitary transformations among the occupied
orbitals, as is the case with traditional CEPA methodologies.  Perhaps even
more unfortunately, this dependence also extends to the virtual space.  The
dependence upon the choice of the virtual orbitals is a consequence of the
symmetry properties
of the topological factor in the occupied and virtual spaces; EPV diagrams
are removed from not only the occupied space, as is the case in CEPA, but from
the virtual space as well.  Figure \ref{FIG:INVAR2} illustrates the deviation
of the energy obtained from calculations with localized orbitals from those
performed using canonical Hartree-Fock orbitals for the H-F bond stretch in
(a) cc-pVDZ and (b) cc-pVTZ basis sets.
We present results for two cases within each basis set: (i) calculations in
which only the occupied orbitals are localized and (ii) calculations in which
we localize both the occupied and virtual orbitals separately.  Clearly, the
2-RDM method is not invariant to unitary transformations in either the occupied
or virtual subspaces.  Importantly, the method's variance with respect to the
occupied space is nearly constant at all bond lengths and across basis sets.
We note that the sign of the percent change for local occupied orbitals changes
between basis sets, meaning that the energy with localized orbitals was lower
than the canonical case in the cc-pVDZ basis but higher in the cc-pVTZ basis.
While the relative magnitudes of the changes are very similar, the change in
sign implies that we cannot assume a systematic change in energy for arbitrary
systems and basis sets when utilizing localized versus delocalized canonical
orbitals. The variance with respect to the virtual space, while being fairly
uniform across basis sets, is highly dependent upon the geometry of the system.
Local virtual orbitals are in general very difficult to determine; minima
are often local in nature in the localization function, and the orbitals
themselves may not vary smoothly with nuclear coordinates.  It is unclear
whether the strong dependence of the energy on the virtual space is a
consequence of these difficulties or some other systematic deficiency
inherent to the p-2RDM$^\prime$/$n$ methods.

\begin{figure}[!htpb]
  \label{FIG:HF}
  \caption
          {Energy per HF molecule as a function of the number of HF molecule
           for a system of non-interacting HF molecules.  The 2-RDM method is
           only rigorously size-extensive when using a basis of localize
           occupied and virtual molecular orbitals.}
  \begin{center}
    \resizebox{11cm}{!}{\includegraphics{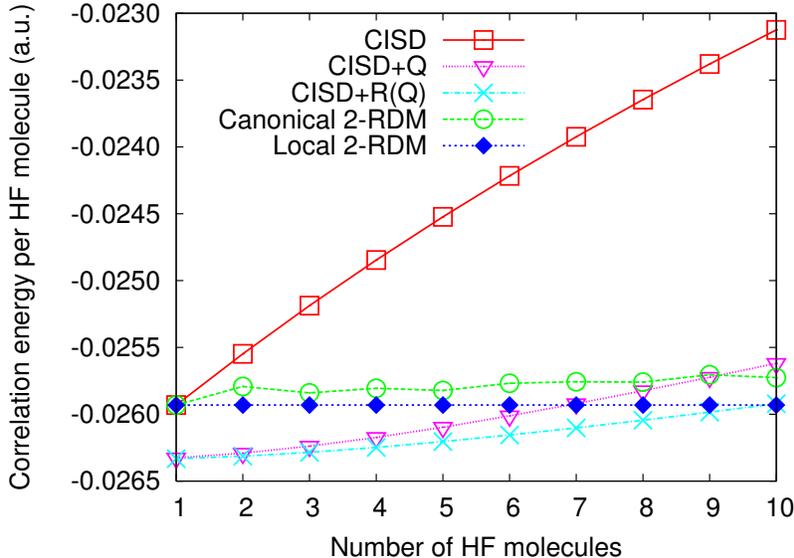}}
  \end{center}
\end{figure}

We further investigate the dependence of the 2-RDM energy with respect to
variations in the virtual space with a size-extensivity example similar to the
infinitely separated two-electron systems treated above; we here investigate
the size-extensivity of the 2-RDM method of infinitely separated
two-{\em hole} systems.  As in the He example above, a size-extensive
double-excitation theory should yield the exact result regardless of system
size.  Figure \ref{FIG:HF}
illustrates the correlation energy for a system of non-interacting HF molecules
in a minimal STO-6G basis set.  The molecules lie parallel to one another in
a line with a distance between each center of mass of 1000 \AA.  The H-F
bond length is taken as the experimentally determined value given in the
computational chemistry comparison and benchmark database (CCCBD)
\cite{REF:7_CCCBD}. The 2-RDM method is utilized with either canonical or
separately localized occupied and virtual orbitals.  As expected, CISD is not
size-extensive.  The 2-RDM method yields effectively size-extensive results
for both choices of orbitals, but the only exactly size-extensive choice
is that in which both the local and virtual orbital spaces are localized.
Both the Davidson and renormalized Davidson-corrected energies exhibit a
strong dependence on system size and are thus not rigorously size-extensive.

\section{Conclusions}

Parametric variational $2$-RDM methods are an accurate and efficient alternative
to traditional {\em ab initio} methods.  They formally scale the same as
CI with single and double excitations and may be implemented at a cost that
is slightly less than coupled cluster with single and double substitutions.
Provided calculations are performed in a local orbital basis, the obtained
energies are rigorously size-extensive. The methods have been previously been
generalized for geometry optimizations, harmonic frequency
analysis, the treatment of open-shell systems, and local correlation
approximations with much success.  Novel parametrizations, unique from that
originally proposed by Kollmar or discussed in this paper have been presented
that result in accuracies similar CCSD(T).

The parametric 2-RDM approach represents a fairly young class of methods, and as
such its relationship to other methods has remained to this point unexplored.
For this reason, we have drawn connections between parametric 2-RDM
methodologies and existing coupled electron pair approximation (CEPA) theories.
We have derived a set of topological factors that correspond to the CEPA/$n$
($n=1,2,3$) family, and the resulting class of methods is a density-based
generalization of CEPA/$n$ called p-2RDM/$n$. Extensive numerical studies of
equilibrium energies, geometries,
and harmonic frequencies have shown that the p-2RDM methods perform very
similarly to their CEPA analogues for a variety of closed-shell systems.
New topological factors have been derived specifically to account for the
exclusion-principle-violating (EPV) terms that arise in the virtual space that
are ignored by standard CEPA methodologies.  Malrieu and coworkers
\cite{REF:7_CEPA9} understood the importance of the balance between occupied and
virtual EPV diagrams in their self-consistent, size-consistent truncated CI
( (SC)$^2$-CISD );  their proposed method is most similar in spirit to the
p-2RDM$^\prime$/$3$ variant discussed herein. Another factor, denoted
p-2RDM$^\prime$/1, is in fact identical to that proposed by Kollmar
\cite{REF:7_Kollmar}. The proper treatment of the virtual space EPV diagrams is
necessary in some situations to obtain physically meaningful results,  as was
demonstrated for the potential energy surface for the CH radical.  Consideration
of the virtual space EPV diagrams for the p-2RDM/3 method is necessary to avoid
an unphysical hump in the dissociation curve. Aside from these numerical
arguments, properly treating
virtual space EPV diagrams is absolutely essential from the standpoint of
density matrix theory in that they are required for a balanced treatment of
particles and holes.  The parametric 2-RDM framework for p-2RDM is also quite
convenient as compared to the standard overlap equation formulation of
traditional CEPA methodologies; the energies obtained from any of the p-2RDM or
p-2RDM$^\prime$ variants presented herein
are stationary points, facilitating the evaluation of geometric derivatives
and the direct computation of density matrices and their associated one- and
two-electron properties.

We have discussed in detail the orbital invariance properties of the CEPA and
p-2RDM methods citing various numerical examples.  The 2-RDM method is indeed
exactly size-extensive, but this claim holds true only under the condition
that one utilizes a basis of localized molecular orbitals.  As such, the
2-RDM method (and thus the p-2RDM and p-2RMD$^\prime$ variants) are not
rigorously invariant to unitary transformations within orbital subspaces.  The
p-2RDM methods display a dependence on the choice of orbitals for the occupied
space while the p-2RMD$^\prime$ methods also vary with the choice of the virtual
orbitals.  One may circumvent any ambiguities with respect to the definition
of the orbital space by always utilizing a basis of local orbitals.  Determining
these orbitals in the occupied subspace is trivial by the Boys localization
criterion \cite{REF:7_Boys}, but may prove problematic for the virtual space
where local orbitals may not be unique and do not necessarily vary smoothly
with nuclear coordinates.  Fortunately,
the variance in the correlation energy with the occupied orbitals is only
a fraction of a percent and at worst on the order of one percent for the virtual
orbitals.  These discrepancies are very small when compared to the percentage
of correlation energy that is not recovered for any of the standard
{\it ab initio} methods with respect to the exact full CI results.

\begin{acknowledgments}

D.A.M. gratefully acknowledges the NSF, the Henry-Camille Dreyfus
Foundation, the David-Lucile Packard Foundation, and the Microsoft
Corporation for their support. A.E.D. acknowledges funding provided
by the Computational Postdoctoral Fellowship through the Computing,
Engineering, and Life Sciences Division of Argonne National
Laboratory.

\end{acknowledgments}


\begin{thebibliography}{99}
\addcontentsline{toc}{section}{10.5   \hspace{0.75mm}References}
\bibitem{REF:7_Challenge} A. J. Coleman and V. I. Yukalov, {\em Reduced Density
                    Matrices: Coulson's Challenge} (Springer-Verlag, New York,
                    2000).
\bibitem{REF:7_RDMBook} D. A. Mazziotti (Ed.), {\em Reduced-Density-Matrix Mechanics: With Application
to Many-electron Atoms and Molecules}, Advances in Chemical Physics, Vol. 134, (Wiley, New York, 2007).
\bibitem{REF:7_NREP1} A. J. Coleman, Rev. Mod. Phys. {\bf 35}, 668 (1962).
\bibitem{REF:7_NREP2} C. Garrod and J. Percus, J. Math. Phys. {\bf 5}, 1756 (1964).
\bibitem{REF:7_VAR1} D. A. Mazziotti, Phys. Rev. Lett. {\bf 93}, 213001 (2004).
\bibitem{REF:7_VAR2} D. A. Mazziotti, J. Chem. Phys. {\bf 121}, 10957 (2004).
\bibitem{REF:7_VAR3} D. A. Mazziotti, Phys. Rev. A {\bf 74}, 032501 (2006).
\bibitem{REF:7_VAR4} D. A. Mazziotti, Acc. Chem. Res. {\bf 39}, 207 (2006).
\bibitem{REF:7_VAR5} G. Gidofalvi and D. A. Mazziotti, J. Chem. Phys. {\bf 126}, 024105 (2007).
\bibitem{REF:7_VAR6} Z. Zhao, B. J. Braams, H. Fukuda, M. L. Overton, and J. K. Percus, J. Chem. Phys {\bf 120}, 2095 (2004).
\bibitem{REF:7_VAR7} E. Canc$\grave{\rm e}$s, G. Stoltz, and M. J. Lewin, J. Chem. Phys. {\bf 125}, 064101 (2006).
\bibitem{REF:7_ACSE1} D. A. Mazziotti, Phys. Rev. Lett. {\bf 97}, 143002 (2006).
\bibitem{REF:7_ACSE2} D. A. Mazziotti, Phys. Rev. A {\bf 75}, 022505 (2007).
\bibitem{REF:7_ACSE3} D. A. Mazziotti, J. Phys. Chem. A {\bf 111}, 12635 (2007).
\bibitem{REF:7_ACSE4} D. A. Mazziotti, J. Chem. Phys. {\bf 126}, 184101 (2007).
\bibitem{REF:7_ACSE5} D. A. Mazziotti, Phys. Rev. A {\bf 76}, 052502 (2007).
\bibitem{REF:7_ACSE6} D. A. Mazziotti, J. Phys. Chem. A {\bf 112}, 13684 (2008).
\bibitem{REF:7_ACSE7} C. Valdemoro, L. M. Tel, D. R. Alcoba, and E. P$\acute{\rm e}$rez-Romero, Theor. Chem. Acc. {\bf 118}, 503 (2007).
\bibitem{REF:7_ACSE8} C. Valdemoro, L. M. Tel, E. E. P$\acute{\rm e}$rez-Romero, and D. R. Alcoba, Int. J. Quantum Chem. {\bf 108}, 1090 (2008).
\bibitem{REF:7_ACSE9} D. A. Mazziotti, Phys. Rev. A {\bf 57}, 4219 (1998).
\bibitem{REF:7_ACSE10} F. Colmenero and C. Valdemoro, Phys. Rev. A {\bf 47}, 979 (1993).
\bibitem{REF:7_ACSE11} H. Nakatsuji and K. Yasuda, Phys. Rev. Lett. {\bf 76}, 1039 (1996).
\bibitem{REF:7_Kollmar} C. Kollmar, J. Chem. Phys. {\bf 125}, 084108 (2006).
\bibitem{REF:7_2-RDMF1} A. E. DePrince III and D. A. Mazziotti, Phys. Rev. A {\bf 73}, 042501 (2007).
\bibitem{REF:7_2-RDMF2} A. E. DePrince III, E. Kamarchik, and D.A. Mazziotti, J. Chem. Phys. {\bf 128}, 234103 (2008).
\bibitem{REF:7_2-RDMF3} A. E. DePrince III and D. A. Mazziotti, J. Phys. Chem. B {\bf 112}, 16158 (2008).
\bibitem{REF:7_PRL} D. A. Mazziotti, Phys. Rev. Lett. {\bf 101}, 253002 (2008); Phys. Rev. A {\bf 81}, 062515 (2010).
\bibitem{REF:7_2-RDMF4} A. E. DePrince III and D. A. Mazziotti, J. Chem. Phys. {\bf 130}, 164109 (2009).
\bibitem{REF:7_2-RDMF5} A. E. DePrince III and D. A. Mazziotti, J. Chem. Phys. {\bf 132}, 034110 (2010).
\bibitem{REF:7_2-RDMF6} A. E. DePrince III and D. A. Mazziotti, J. Chem. Phys. {\bf 133}, 034112 (2010).

\bibitem{REF:7_CEPA1} S. Koch and W. Kutzelnigg, Theoret. Chim. Acta {\bf 59}, 387 (1981).
\bibitem{REF:7_CEPA2} R. Ahlrichs, Comput. Phys. Comm. {\bf 17}, 31 (1979).
\bibitem{REF:7_CEPA3} W. Meyer, Int. J. Quant. Chem. {\bf 5}, 341 (1971).
\bibitem{REF:7_CEPA4} W. Meyer, J. Chem. Phys. {\bf 58}, 1017 (1973).
\bibitem{REF:7_CEPA5} P. Taylor, G. B. Bacskay, and N. S. Hush, Chem. Phys. Lett. {\bf 41}, 444 (1976).
\bibitem{REF:7_CPF1} R. Ahlrichs, P. Scharf, and C. Ehrhardt, J. Chem. Phys. {\bf 82}, 890 (1985).
\bibitem{REF:7_CPF2} R. J. Gdanitz and R. Ahlrichs, Chem. Phys. Lett. {\bf 143}, 413 (1988).
\bibitem{REF:7_CEPA6} F. Wennmohs, F. Neese, Chem. Phys. {\bf 343}, 217 (2008).
\bibitem{REF:7_CEPA7} M. Nooijen and R. J. Le Roy, J. Molec. Struct.: THEOCHEM {\bf 768}, 25 (2006).
\bibitem{REF:7_CEPA8} F. Neese, A. Hansen, F. Wennmohs, and S. Grimme, Acc. Chem. Res. {\bf 42}, 641 (2009).
\bibitem{REF:7_CEPA9} J.-P. Daudey, J.-L. Heully, and J.-P. Malrieu, J. Chem. Phys. {\bf 99}, 1240 (1993).

\bibitem{ED09} A. E. DePrince III, {\em A Parametric Approach to Variational Two-electron Reduced Density Matrix Theory}, Ph.D. thesis, Department of Chemistry, The University of Chicago, 2009.
\bibitem{NK10} C. Kollmar and F. Neese, Mol. Phys. (2010), DOI: 10.1080/00268976.2010.496743.

\bibitem{REF:7_CEPA10} J.-P. Malrieu,H. Zhang, and J. Ma, Chem. Phys. Lett. In Press, Corrected Proof (2010).
\bibitem{REF:7_CEPA11} C. Kollmar and A. He$\beta$elmann, Theor. Chem. Acc. (in press 2010).
\bibitem{REF:7_MOLPRO}{\small MOLPRO}, version 2006.1, a package of {\em ab initio} programs, designed by H.-J. Werner, P. J. Knowles, R. Lindh, F. R. Manby, M. Sch\"{u}tz {\em et al.} (see http://www.molpro.net).
\bibitem{REF:7_PSI3}  T. D. Crawford, C. D. Sherrill, E. F. Valeev {\it et al.}, J. Comput. Chem. {\bf 28}, 1610 (2007).
\bibitem{REF:7_GAMESS} M. W. Schmidt, K. K. Baldridge, J. A. Boatz, S. T. Elbert, M. S. Gordon, J. H. Jensen, S. Koseki, N. Matsunaga, K. A. Nguyen, S. Su, T. L. Windus, M. Dupuis, and J. A. Montgomery Jr, J. Comput. Chem. {\bf 14}, 1347 (1993).
\bibitem{REF:7_BART_INVAR} G. D. Purvis and R. J. Bartlett, J. Chem. Phys. {\bf 68}, 2114 (1978).
\bibitem{REF:7_ADZ} A. Schafer, H. Horn, and R. Ahlrichs, J. Chem. Phys. {\bf 97}, 2571 (1992).
\bibitem{REF:7_Boys}S. F. Boys, Rev. Mod. Phys. {\bf 32} 296 (1960).
\bibitem{REF:7_Davidson} S. R. Langhoff and E. R. Davidson, Int. J. Quant. Chem. {\bf 8}, 61 (1974).
\bibitem{REF:7_RD1} E. M. Siegbahn, Chem. Phys. Lett. {\bf 55}, 386 (1978).
\bibitem{REF:7_RD2} L. Meissner, Chem. Phys. Lett. {\bf 146}, 204 (1988).
\bibitem{REF:7_CCCBD} Computational Chemistry Comparison and Benchmark Database, http://srdata.nist.gov/cccbdb

\end{thebibliography}
\end{document}